\documentclass[prl,aps,superscriptaddress,twocolumn,amssymb,amsmath,longbibliography]{revtex4-2}
\usepackage{graphicx}
\usepackage{theorem}
\usepackage{dcolumn}
\usepackage{bm}
\usepackage{mathrsfs}
\usepackage{setspace}

\begin{document}
\title{Turbulent diffusion and dispersion in a superfluid}

\author{Yuan Tang}
\affiliation{National High Magnetic Field Laboratory, 1800 East Paul Dirac Drive, Tallahassee, Florida 32310, USA}
\affiliation{Mechanical Engineering Department, FAMU-FSU College of Engineering, Florida State University, Tallahassee, Florida 32310, USA}

\author{Sosuke Inui}
\affiliation{National High Magnetic Field Laboratory, 1800 East Paul Dirac Drive, Tallahassee, Florida 32310, USA}
\affiliation{Mechanical Engineering Department, FAMU-FSU College of Engineering, Florida State University, Tallahassee, Florida 32310, USA}

\author{Yiming Xing}
\affiliation{National High Magnetic Field Laboratory, 1800 East Paul Dirac Drive, Tallahassee, Florida 32310, USA}
\affiliation{Mechanical Engineering Department, FAMU-FSU College of Engineering, Florida State University, Tallahassee, Florida 32310, USA}

\author{Yinghe Qi}
\affiliation{National High Magnetic Field Laboratory, 1800 East Paul Dirac Drive, Tallahassee, Florida 32310, USA}
\affiliation{Mechanical Engineering Department, FAMU-FSU College of Engineering, Florida State University, Tallahassee, Florida 32310, USA}

\author{Wei Guo}
\email[Corresponding author: ]{wguo@eng.famu.fsu.edu}
\affiliation{National High Magnetic Field Laboratory, 1800 East Paul Dirac Drive, Tallahassee, Florida 32310, USA}
\affiliation{Mechanical Engineering Department, FAMU-FSU College of Engineering, Florida State University, Tallahassee, Florida 32310, USA}

\date{\today}

\begin{abstract}
Single-body diffusion and two-body dispersion are fundamental processes in classical turbulence, governing particle mixing and transport. However, their behaviors in superfluid turbulence remain largely unexplored. In this study, we numerically investigate the diffusion and relative dispersion of quantized vortices and superfluid parcels in the 0 K limit in two distinct turbulence regimes: ultra-quantum turbulence, characterized by a randomized vortex tangle, and quasiclassical turbulence, in which locally polarized vortices create large-scale flows resembling classical turbulence. Our results reveal that while vortex segments exhibit similar superdiffusion behavior at short times in both regimes, superfluid parcels behave differently—following the same superdiffusion scaling in ultra-quantum turbulence but deviating significantly in quasiclassical turbulence. This contrast provides a key clue to the origin of short-time superdiffusion, a puzzle since its recent discovery. Additionally, we show that two-body dispersion of both vortex segments and superfluid parcels exhibits distinct scaling behaviors in ultra-quantum and quasiclassical turbulence, highlighting fundamental differences in these two turbulence regimes. Our findings bridge a critical gap in superfluid turbulence research, offering new insights into turbulent transport in inviscid quantum fluids.

\end{abstract}
\maketitle

Turbulent diffusion and dispersion are fundamental to understanding transport phenomena in fluid dynamics~\cite{Sreeni-2019-PNAS}. In classical turbulence, the chaotic motion of eddies governs how individual fluid parcels spread over time (single-body diffusion) and how initially close pairs of parcels separate (two-body dispersion), both of which follow well-characterized scaling laws. For instance, for a fluid parcel with position vector $\mathbf{r}(t)$, its mean square displacement (MSD), defined as $\langle \Delta \mathbf{r}^2(t) \rangle = \langle |\mathbf{r}(t)-\mathbf{r}(0)|^2 \rangle$, follows a ballistic scaling, $\langle \Delta \mathbf{r}^2(t) \rangle \approx u_{rms}^2 t^2$, at short times ($t \ll \tau_L$) and transitions to a diffusive scaling, $\langle \Delta \mathbf{r}^2(t) \rangle = 2\Gamma_T t$, at long times ($t \gg \tau_L$) \cite{Davidson-2004}. Here, $u_{rms}$ is the root-mean-square (rms) velocity, $\tau_L = L/u_{rms}$ is the integral timescale with $L$ being the integral length scale, and $\Gamma_T = u_{rms}^2 \tau_L$ is the turbulent eddy diffusivity \cite{Pope-2000}. Meanwhile, in fully developed turbulence, the mean square separation $\langle \mathbf{\Delta}^2(t) \rangle = \langle |\mathbf{r}_1(t) - \mathbf{r}_2(t)|^2 \rangle$ of two initially close fluid parcels follows Richardson–Obukhov $t^3$ law \cite{Richardson-1926,Batchelor-1952}, $\langle \mathbf{\Delta}^2(t) \rangle = g \epsilon t^3$, for times $\tau_0<t<\tau_L$, when the parcel separation remains within the inertial range. Here, $\epsilon$ is the turbulent energy dissipation rate, $g\simeq0.5$ is the Richardson constant, and $\tau_0 = |\mathbf{\Delta}(0)|^{2/3} / \epsilon^{1/3}$ is the memory time \cite{Hunt-Vassilicos-1991}. Extensive studies have explored these scaling laws in classical fluids \cite{Sawford-2001,Boffetta-2002-POF,Toschi-2009,Salazar-2009-ARFM,Schumacher-2014,Bragg-2016,Buaria-2019,Tan-2022}.

In an invcisd superfluid, the situation becomes intriguing. Turbulence in a superfluid can arise from a chaotic tangle of quantized vortex lines \cite{Vinen-2002-JLP}, which are topological defects characterized by a quantized circulation $\kappa=h/m$, where $h$ is Plank's constant and $m$ is the mass of the bosons constituting the superfluid~\cite{Tilley-1990-book}. The vortices evolve with time and can reconnect when they intersect~\cite{Koplik-1993-PRL}. Depending on the structure of the vortex tangle, superfluid turbulence can manifest in two distinct regimes~\cite{Volovik-2003-JETP,Walmsley-2014-PNAS}. The first, known as quasiclassical turbulence (QCT), emerges when the vortices locally align and form bundles that mimic classical vortices~\cite{Baggaley-2012-PRL}. In this regime, the induced velocity field exhibits classical behavior at length scales larger than the mean inter-vortex spacing $\ell$~\cite{Tang-2020-PRF,Tang-2021-PRB}. In contrast, when the vortices are arranged randomly, it gives rise to ultra-quantum turbulence (UQT), a regime with no classical counterpart, where the flow field fluctuates at scales comparable to $\ell$ without large-scale motion \cite{Walmsley-2008-PRL}. Both QCT and UQT can decay even at the 0 K limit without a viscous thermal component. At scales comparable to $\ell$, vortex reconnections excite Kelvin waves on the vortices \cite{Fonda-2014-PNAS}. These waves then cascade energy to smaller scales through nonlinear interactions, eventually causing quasiparticle emissions \cite{Kozik-2004-PRL}. While extensive studies have examined the energy cascade and spectral properties in these turbulence regimes \cite{Vinen-2002-JLP, Kozik-2004-PRL,Lvov-2007-PRB, Boue-2012-PRE,Gao-2017-PRB}, much less is known on how vortices and superfluid parcels diffuse and disperse.

%While extensive studies have examined the energy cascade and spectral properties in these turbulence regimes, much less is known on how vortex segments and superfluid parcels diffuse and disperse.
%In the 0 K limit, where viscous dissipation is absent, superfluid turbulence dissipates at scales comparable to $\ell$, driven by vortex reconnections that trigger quasiparticle emissions and generate Kelvin waves on the vortices. These waves facilitate an energy cascade to scales approaching the vortex core size through nonlinear interactions, ultimately resulting in further quasiparticle emissions. While extensive studies have examined the energy cascade and spectral properties in these regimes, far less is known about the diffusion and dispersion of vortex segments and superfluid parcels.

\begin{figure*}[t]
\centering
\includegraphics[width=1\linewidth]{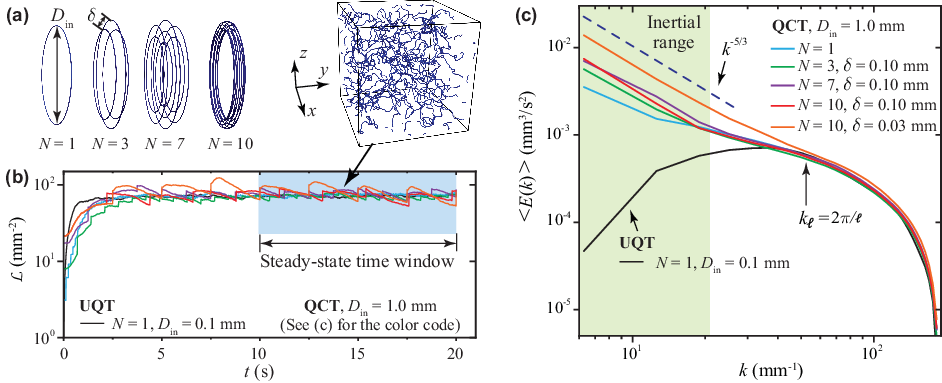}
\caption{(a) Schematic showing the configurations of the injected vortex rings and ring bundles. (b) Time evolution of the vortex-line density $\mathcal{L}$. A snapshot of the vortex tangle for the case $D_{\text{in}}=1$ mm, $N=10$, and $\delta=0.03$ mm is included. (c) Ensemble-averaged turbulence energy spectrum over the steady-state time window for various injection conditions.}
\label{Fig1}
\end{figure*}

Some recent experiments reported the observations of superdiffusion of quantized vortices in UQT, where the MSD follows $\langle \Delta \mathbf{r}^2(t) \rangle \propto t^\gamma$ at short times with an exponent $\gamma\simeq1.6$, transitioning to normal diffusion with $\gamma\simeq2$ for $t\gg\tau_{\ell}$, where $\tau_{\ell}=\ell/u_{rms}$ is the characteristic time a vortex segment moves before undergoing reconnections \cite{Tang-2021-PNAS,Boltnev-SR-2023}. Numerical simulations confirmed this scaling behavior and attributed it to an intrinsic temporal correlation of the vortex velocity for $t<\tau_{\ell}$ \cite{Yui-PRL-2022}, though the mechanism underlying this velocity correlation remains a puzzle. It is also unclear whether similar behavior persists in QCT and whether superfluid parcels follow the same diffusion laws. Regarding turbulent dispersion, apart from a preliminary study on vortex dispersion in two-dimensional UQT \cite{Poole-JLTP-2004}, a comprehensive understanding of possible dispersion scaling laws in both QCT and UQT is still lacking. In this work, we conduct a systematic numerical study on the diffusion and dispersion of quantized vortices and superfluid parcels in both turbulence regimes at the 0 K limit to uncover the intrinsic transport properties of superfluid turbulence. This study not only advances our understanding of superfluid turbulence but also provides useful implications for practical applications, such as particle dynamics and spectroscopy in superfluid $^4$He (He II) droplets \cite{Lehmann1998,Callegari2001,Toennies2004,Vilesov2018} and vortex-assisted nanowire synthesis in He II \cite{Moroshkin2010,Gordon2015,Gordon2017}.

\emph{Generation of UQT and QCT.}—To study vortex dynamics, we adopt the vortex filament model \cite{Schwarz-1988-PRB}, where each vortex line is represented as a zero-thickness filament discretized into a series of points. To generate UQT, we use the method adopted in Ref.~\cite{Yui-PRL-2022}, where randomly oriented small vortex rings (diameter $D_{\text{in}}=0.1$ mm) are repetitively injected at random locations within a cubical computational box (side length: $D=1$ mm) with periodic boundary conditions in all three directions. The injection occurs at fixed intervals of $t_{\text{in}}$ to facilitate the buildup of a vortex tangle. In contrast, generating QCT requires energy injection at large length scales \cite{Walmsley-2008-PRL}. In this work, we tested injecting individual vortex rings with $D_{\text{in}}=1$ mm as well as vortex ring bundles consisting of $N=3$, 7, or 10 co-axial rings arranged in a triangular lattice within the plane containing the central axis. The bundles have an averaged $D_{\text{in}}=1$ mm and an inter-ring distance of $\delta=0.1$ mm or 0.03 mm, as illustrated in Fig.~\ref{Fig1}(a). A filament at $\mathbf{s}$ moves at the local superfluid velocity $\mathbf{u}_s$ given by the Biot-Savart law\cite{Schwarz-1988-PRB,Tsubota-2000-PRB}:
\begin{equation}
\frac{d\mathbf{s}}{dt}=\mathbf{u}_s(\mathbf{s},t)=\frac{\kappa}{4\pi}\int\frac{(\mathbf{s_1}-\mathbf{s})\times d\mathbf{s_1}}{|\mathbf{s_1}-\mathbf{s}|^3}.
\label{Eq1}
\end{equation}
where the integration is performed along all the vortices. Since the integrant diverges at $\mathbf{s}$, we follow Adachi \emph{et al.} and calculate the integral as the sum of the local contribution and the non-local contribution \cite{Adachi-2010-PRB}:
\begin{equation}
\frac{d\mathbf{s}}{dt}=\beta_l\mathbf{s}'\times\mathbf{s}''+\frac{\kappa}{4\pi}\int_{\mathbf{s_1}\neq\mathbf{s}}\frac{(\mathbf{s_1}-\mathbf{s})\times d\mathbf{s_1}}{|\mathbf{s_1}-\mathbf{s}|^3},
\end{equation}
where the prime denotes derivative with respect to the arc length of the vortex filament at $\mathbf{s}$, and $\beta_l$ is a vortex core-size dependant coefficient \cite{Schwarz-1988-PRB} (see Supplemental Materials \cite{Supplemental}, which includes Refs.~\cite{Press-1992-book, Baggaley-2012-JLTP,Yurkina-2021-LTP}). When two vortices approach to have a separation less than our spatial resolution $\Delta\xi_{min}=0.008$ mm, we reconnect them at the location of the minimum separation, following the procedures as detailed in Refs.~\cite{Tsubota-2000-PRB,Baggaley-2012-JLTP}. We also remove small vortex loops with lengths less than $5\Delta\xi_{min}$ to account for the cascade loss of the vortices \cite{Tsubota-2000-PRB}. Fig.~\ref{Fig1}(b) shows the time evolution of the vortex-line density $\mathcal{L}$ (defined as the total vortex length per unit volume) for various injection conditions, where $\kappa=10^{-3}$ cm$^2$/s for He II is used. After an initial transient of 2–3 s, $\mathcal{L}$ settles to a steady level. To ensure a fair comparison, we adjust $t_{\text{in}}$ for each case so that the steady-state $\mathcal{L}$ remains about 75 mm$^{-2}$ for all cases, corresponding to a mean inter-vortex spacing of $\ell=\mathcal{L}^{-1/2}\simeq 0.11$ mm. A representative snapshot of the vortex tangle for the case with $D_{\text{in}}=1$ mm, $N=10$, and $\delta=0.03$ mm is also shown in Fig.~\ref{Fig1}(b).

\begin{figure*}[t]
\centering
\includegraphics[width=1\linewidth]{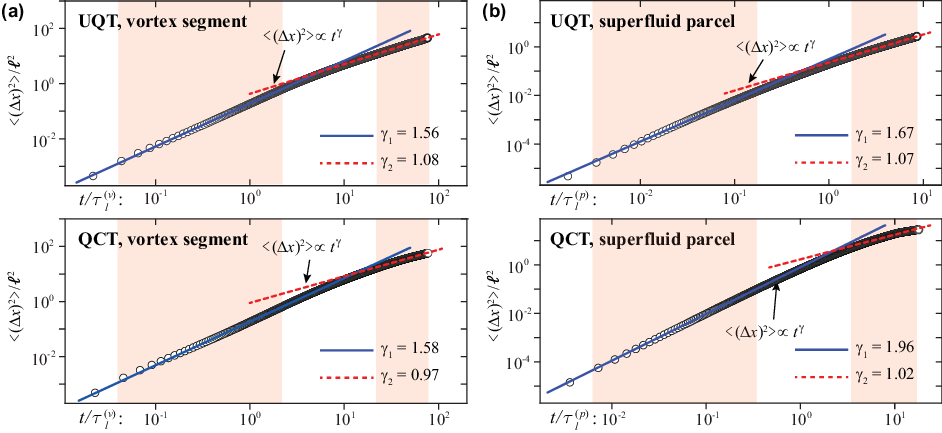}
\caption{Mean square displacement (MSD), $\langle\Delta x^2(t)\rangle$, along the $x$-direction for (a) vortex segments and (b) superfluid parcels in UQT (top) and QCT (bottom). Solid and dashed lines represent power-law fits to the data within the shaded regions.}
\label{Fig2}
\end{figure*}

%(a) Mean square displacement (MSD) of vortex segments $\langle\Delta x^2(t)\rangle$ along the $x$-direction for UQT (upper) and QCT (lower). (b) MSD of superfluid parcels along the $x$-direction for UQT (upper) and QCT (lower). The solid and dashed lines are power-law fits to the data in the shaded regions.

Knowing the vortex tangle configuration at time $t$, the energy spectrum $E(k,t)$ as a function of wavenumber $k$, associated with the superfluid velocity field, can be computed using the expression from Ref. \cite{Yurkina-2021-LTP} (see Supplemental Materials \cite{Supplemental}). This spectrum is then ensemble-averaged over the steady-state time window to obtain $\langle E(k)\rangle$. Results for various injection conditions are shown in Fig.~\ref{Fig1}(c). For UQT, where the energy is injected at a scale comparable to $\ell$, the spectrum $\langle E(k)\rangle$ peaks near $k_{\ell}=2\pi/\ell$, as expected \cite{Walmsley-2008-PRL}. For QCT, we observe that in the low-$k$ regime, $\langle E(k)\rangle$ increasingly approaches the classical Kolmogorov scaling $k^{-5/3}$ as bundles of more closely packed rings are used for injection. In contrast, injecting randomly oriented individual rings at large scales fails to deliver sufficient energy density to establish an clear inertial range, a key consideration for those aiming to generate QCT via ring injection. We also fit the spectrum using $\langle E(k)\rangle=\alpha \epsilon^{2/3}k^{-5/3}$ for the case with $D_{\text{in}}=1$ mm, $N=10$, and $\delta=0.03$ mm, where the energy dissipation rate $\epsilon=0.106$ mm$^2$/s$^3$ is determined from the measured energy increase following each bundle injection (see Supplemental Materials \cite{Supplemental}). The fitted Kolmogorov constant is $\alpha=1.43$, which is close to the classical value $\alpha\simeq1.5$ \cite{Sreenivasan-1995-PF}. In what follows, we focus on this case and the UQT case for diffusion and dispersion analysis. Table \ref{Table1} summarizes the rms velocities of vortex segments $u^{(v)}_{rms}$ and superfluid parcels $u^{(p)}_{rms}$, along with the calculated timescales $\tau_{\ell}=\ell/u_{rms}$ for the two cases.

\begin{table}[t]
\centering
\caption{Key flow properties for the UQT and QCT cases.}
\label{Table1}
\begin{tabular}{>{\centering\arraybackslash}p{1cm}
                >{\centering\arraybackslash}p{2.2cm}
                >{\centering\arraybackslash}p{1.1cm}
                >{\centering\arraybackslash}p{2.2cm}
                >{\centering\arraybackslash}p{1.1cm}}
\hline
    & $u^{(v)}_{rms}$ (mm/s) & $\tau^{(v)}_{\ell}$ (s) & $u^{(p)}_{rms}$ (mm/s) & $\tau^{(p)}_{\ell}$ (s) \\
\hline
UQT & 2.57 & 0.046 & 0.21 & 0.56 \\
QCT & 2.50 & 0.044 & 0.41 & 0.27 \\
\hline
\end{tabular}
\end{table}

\emph{Turbulent diffusion in UQT and QCT.}—To study vortex diffusion, we track randomly selected vortex-filament points using the tagging method described in Ref. \cite{Yui-PRL-2022} and analyze their MSD along each axis during the steady-state time window. The resulting vortex MSDs in the $x$ direction $\langle\Delta x^2(t)\rangle=\langle[x(t_0+t)-x(t_0)]^2\rangle$ for both UQT and QCT are shown in Fig.~\ref{Fig2}(a). For UQT, the MSD exhibits superdiffusion scaling $\langle\Delta x^2(t)\rangle \propto t^{\gamma_1}$ with $\gamma_1\simeq1.6$ at short times around and below the vortex reconnection timescale $\tau^{(v)}_{\ell}$. This superdiffusion arises from temporal correlations in vortex velocity \cite{Yui-PRL-2022}. At very short times ($t<10^{-2}\tau^{(v)}_{\ell}$) when the velocity correlation reaches full coherence, ballistic scaling $\langle\Delta x^2(t)\rangle \propto t^2$ can emerge. At longer times, the MSD transitions to normal diffusion, $\langle\Delta x^2(t)\rangle \propto t^{\gamma_2}$ with $\gamma_2\simeq1$, as vortex reconnections disrupt the correlations and randomize vortex motion \cite{Yui-PRL-2022}. Interestingly, the vortex MSD in QCT shows very similar behavior despite the fundamental differences between the two turbulence regimes, suggesting a possibly universal mechanism controlling vortex diffusion.

\begin{figure*}[t]
\centering
\includegraphics[width=1\linewidth]{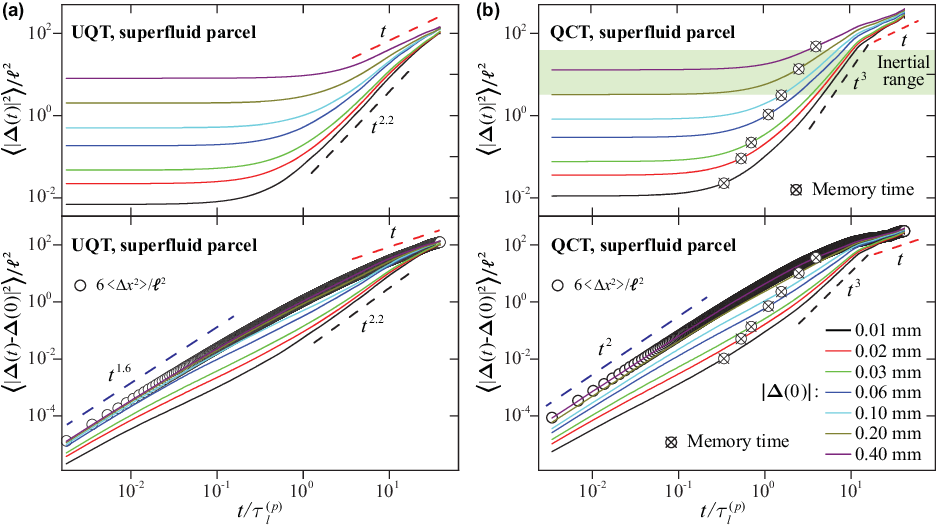}
\caption{Simulated superfluid parcel-pair dispersion in (a) UQT and (b) QCT. The solid curves for different colors correspond to different initial pair separation $|\mathbf{\Delta}(0)|$ ranging from 0.01 mm to 0.4 mm. Open circles represent $6\langle \Delta x^2(t)\rangle/\ell^2$, obtained from individual-parcel diffusion data. Circles with crosses in (b) mark the memory time $\tau_0$ for each case.}
\label{Fig3}
\end{figure*}

Further insight comes from our analysis of the apparent diffusion of superfluid parcels. We begin by randomly selecting parcels at time $t_0$. Their initial velocities $\mathbf{u}_s(\mathbf{r})$ are computed using Eq.~(\ref{Eq1}) based on the vortex configuration at $t_0$. The parcels are then advected to new positions $\mathbf{r}(t_0+\Delta t)=\mathbf{r}(t_0)+\mathbf{u}_s(\mathbf{r})\cdot \Delta t$, and their velocities are updated using the vortex configuration at $t_0+\Delta t$. Repeating this procedure yields complete parcel trajectories, enabling evaluation of their ensemble-averaged MSDs along all three axes. Fig.~\ref{Fig2}(b) shows the MSDs in the $x$ direction, $\langle\Delta x^2(t)\rangle$, for both UQT and QCT. In UQT, parcels exhibit the same diffusion scaling as vortex segments, though with smaller amplitude in $\langle\Delta x^2(t)\rangle$. Interestingly, in QCT, the parcel MSD displays ballistic scaling $\langle\Delta x^2(t)\rangle \propto t^{\gamma_1}$ with $\gamma_1\simeq2$ at short times, before transitioning to normal diffusion at longer times—resembling classical turbulent diffusion and in stark contrast to the vortex diffusion in the same QCT.

To understand these observations, we note that while both vortex segments and superfluid parcels follow the superfluid velocity field $\mathbf{u}_s$, a key difference is that vortex segments experience the local induction term $\beta_l\mathbf{s}'\times\mathbf{s}''$, which is absent for most parcels since they are not typically on the vortices. In UQT, both the local and non-local contributions to $\mathbf{u}_s$ render similar temporal correlations, resulting in comparable superdiffusion scaling for both vortex segments and parcels. In QCT, however, the non-local term generates large-scale eddy motions due to vortex bundling and polarization. As a result, parcels—whose motion is largely governed by the non-local field—are advected by these coherent flows, leading to fully correlated velocities at short times and thus ballistic diffusion. In contrast, vortex segments in QCT remain dominated by the local induction term, which yields only power-law temporal velocity correlations and preserves the superdiffusion scaling seen in UQT. This limited correlation arises because the vortices have local curvature radii comparable to $\ell$ \cite{Vinen-2002-JLP}, causing them to rotate and move in a way that prevents a full temporal correlation. To support this interpretation, we performed simulations using the local induction approximation (LIA), where vortices are evolved solely by the local term, as originally adopted by Schwarz \cite{Schwarz-1988-PRB}. As shown in the Supplemental Materials \cite{Supplemental}, the vortex MSDs in both UQT and QCT under LIA show similar short-time superdiffusion as in Fig.~\ref{Fig2}(a), confirming that this behavior is governed by the local induction effect. %This finding may motivate future analytical derivations of the observed scaling $\langle\Delta x^2(t)\rangle\propto t^{1.6}$ in the LIA framework.

\emph{Turbulent dispersion in UQT and QCT.}—To analyze two-body dispersion, we track randomly selected pairs of vortex segments and superfluid parcels, measuring their separation over time as $|\mathbf{\Delta}(t)|=|\mathbf{r}_1(t)-\mathbf{r}_2(t)|$. The mean square separation $\langle|\mathbf{\Delta}(t)|^2\rangle$ is then ensemble-averaged over many pairs within the steady-state window. Results for parcel pairs with various initial separations $|\mathbf{\Delta}(0)|$ are shown in the upper panels of Fig. \ref{Fig3}(a) and (b) for UQT and QCT, respectively. Similar to classical fluids \cite{Sawford-2001,Boffetta-2002-POF,Toschi-2009}, an initial ``frozen'' regime is observed, where $\langle|\mathbf{\Delta}(t)|^2\rangle$ remains nearly constant due to minimal relative motion. As turbulent fluctuations accumulate, the separations begin to grow. For QCT, we compute the memory time $\tau_0=|\mathbf{\Delta}(0)|^{2/3}/\epsilon^{1/3}$ for each case. When the initial separation is sufficiently small, we indeed observe the Richardson $t^3$ scaling within the inertial range (highlighted in green in Fig.~\ref{Fig3}(b)) at $t\gg\tau_0$, followed by a transition to normal diffusion. Fitting the classical expression $\langle|\mathbf{\Delta}^2(t)|\rangle=g\epsilon t^3$ to the QCT curve with $|\mathbf{\Delta}(0)|=0.01$ mm yields a Richardson constant of $g\simeq 0.18$, slightly below the classical value of 0.5, likely due to our limited inertial range \cite{Sawford-2008-POF, Borgas-1994-JFM}. In UQT, as $|\mathbf{\Delta}(0)|$ decreases, the curves also converge toward a power-law scaling $\langle|\mathbf{\Delta}(t)|^2\rangle\propto t^{2.2}$, which transitions to normal diffusion at late times. This anomalous $t^{2.2}$ scaling—distinct from classical turbulence—represents a novel observation that may be attributed to differences in large-scale flows between UQT and QCT.

We also analyzed the scaling behavior of $\langle|\mathbf{\Delta}(t)-\mathbf{\Delta}(0)|^2\rangle$ to exclude the influence of the initial ``frozen'' regime. The results for parcel pairs are presented in the lower panels of Fig.~\ref{Fig3}. For pairs with large initial separations, the $\langle|\mathbf{\Delta}(t)-\mathbf{\Delta}(0)|^2\rangle/\ell^2$ curves align closely with the diffusion data $6\langle \Delta x^2(t)\rangle/\ell^2$ for individual parcels. This is expected when the motions of the two parcels are effectively uncorrelated. As $|\mathbf{\Delta}(0)|$ decreases, the curves converge and recover the dispersion scaling behavior discussed earlier. This type of plot, which simultaneously conveys information about single-body diffusion and two-body dispersion, may serve as a useful tool for data presentation.

\begin{figure}[t]
\centering
\includegraphics[width=1\linewidth]{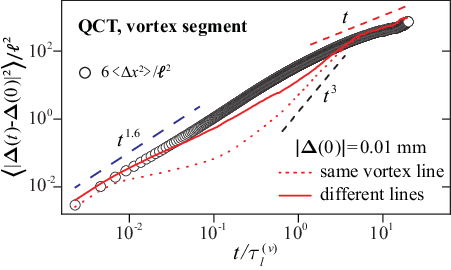}
\caption{Dispersion behavior of vortex segments in QCT with an initial separation of $|\mathbf{\Delta}(0)|=0.01$ mm. The red solid and dotted curves represent segments pairs initially located on the same vortex line and on different vortex lines, respectively.}
\label{Fig4}
\end{figure}

The dispersion behavior of vortex segments closely resembles that of superfluid parcels in both UQT and QCT (see Supplemental Materials \cite{Supplemental}). Furthermore, since quantized vortices have well-defined cores—unlike those in classical fluids—they can be unambiguously tracked for analyses that are difficult to perform in classical turbulence. For instance, we can distinguish vortex-segment pairs that are initially on the same vortex line from those on different vortex lines, and analyze their separation behavior separately. The results for vortex-segment pairs with an initial separation of $|\mathbf{\Delta}(0)|=0.01$ mm are shown in Fig.~\ref{Fig4}. For pairs initially on the same vortex line, the dispersion curve exhibits the scaling behavior discussed earlier. However, for pairs initially on different vortex lines, the dispersion closely follows the single-segment diffusion curve $6\langle \Delta x^2(t)\rangle/\ell^2$, indicating uncorrelated motion between the segments. This observation suggests that the $t^3$ dispersion scaling for vortex segments in QCT is likely associated with stretching of vortices due to large-scale flows.

\emph{Discussion.}—Our work reveals that the short-time superdiffusion scaling of quantized vortices—regardless of the turbulence regime—is governed by the local induction term. This insight may inspire future analytical derivations of the observed scaling $\langle\Delta x^2(t)\rangle \propto t^{1.6}$ within the LIA framework. Additionally, the dispersion scaling observed in UQT, featuring a novel $t^{2.2}$ behavior, points to unexplored dynamics warranting further study. The method of generating idealized QCT via packed-ring bundle injection is also noteworthy. Owing to the well-defined structure of quantized vortices, such QCT simulations enable statistical analyses that are challenging in classical turbulence, offering new opportunities for advancing our understanding of classical turbulent transport through a quantum analogy.

\begin{acknowledgments}
The authors acknowledge the support from the Gordon and Betty Moore Foundation through Grant DOI 10.37807/gbmf11567 and the National Science Foundation (NSF) under Award No. OSI-2426768. The work was conducted at the National High Magnetic Field Laboratory at Florida State University, which is supported by the National Science Foundation Cooperative Agreement No. DMR-2128556 and the state of Florida.
\end{acknowledgments}

\bibliographystyle{apsrev4-2}
\bibliography{Ref-diffusion}

\end{document}

\title{Supplemental Materials for: Turbulent diffusion and dispersion in a superfluid}
\author{Yuan Tang}
\affiliation{National High Magnetic Field Laboratory, 1800 East Paul Dirac Drive, Tallahassee, Florida 32310, USA}
\affiliation{Mechanical Engineering Department, FAMU-FSU College of Engineering, Florida State University, Tallahassee, Florida 32310, USA}

\author{Sosuke Inui}
\affiliation{National High Magnetic Field Laboratory, 1800 East Paul Dirac Drive, Tallahassee, Florida 32310, USA}
\affiliation{Mechanical Engineering Department, FAMU-FSU College of Engineering, Florida State University, Tallahassee, Florida 32310, USA}

\author{Yiming Xing}
\affiliation{National High Magnetic Field Laboratory, 1800 East Paul Dirac Drive, Tallahassee, Florida 32310, USA}
\affiliation{Mechanical Engineering Department, FAMU-FSU College of Engineering, Florida State University, Tallahassee, Florida 32310, USA}

\author{Yinghe Qi}
\affiliation{National High Magnetic Field Laboratory, 1800 East Paul Dirac Drive, Tallahassee, Florida 32310, USA}
\affiliation{Mechanical Engineering Department, FAMU-FSU College of Engineering, Florida State University, Tallahassee, Florida 32310, USA}

\author{Wei Guo}
\email[Corresponding author: ]{wguo@eng.famu.fsu.edu}
\affiliation{National High Magnetic Field Laboratory, 1800 East Paul Dirac Drive, Tallahassee, Florida 32310, USA}
\affiliation{Mechanical Engineering Department, FAMU-FSU College of Engineering, Florida State University, Tallahassee, Florida 32310, USA}

\maketitle

% 1) Numerical method, resolution, dissipation, etc.
% 2) Expression of energy spectrum and energy dissipation rate of QCT.
% 3) LIA diffusion results
% 4) Dispersion scaling of vortex segments, one figure for UQT and QCT with both D and D-D0.
% *) add our own papers

\section{Vortex filament method}
In this work, we employ the vortex filament model to investigate vortex dynamics and the resulting superfluid velocity field~\cite{Schwarz-1988-PRB}. In this framework, each quantized vortex line is represented as a zero-thickness filament discretized into a series of points. In our simulations, the point spacing $\Delta\xi$ ranges from $\Delta\xi_{\text{min}} = 0.008$ mm to $\Delta\xi_{\text{max}} = 0.024$ mm. In the zero-temperature limit, where the viscous thermal component is absent, a filament point at position $\mathbf{s}$ moves with the local superfluid velocity $\mathbf{v}_s(\mathbf{s})$, given by:
\begin{equation}
\frac{d\mathbf{s}}{dt}=\mathbf{v}_s(\mathbf{s})=\mathbf{v}_0(\mathbf{s})+\mathbf{v}_{in}(\mathbf{s}),
\label{Eq-S1}
\end{equation}
where $\mathbf{v}_0(\mathbf{s})$ is the imposed background superfluid velocity, and $\mathbf{v}_{in}(\mathbf{s})$ denotes the velocity induced at $\mathbf{s}$ by all vortex filaments, as given by the Biot–Savart law~\cite{Schwarz-1988-PRB}:
\begin{equation}
\mathbf{v}_{in}(\mathbf{s})=\frac{\kappa}{4\pi}\int\frac{(\mathbf{s_1}-\mathbf{s})\times d\mathbf{s_1}}{|\mathbf{s_1}-\mathbf{s}|^3},
\end{equation}
where the integration is carried out along all the vortex filaments. Since this integrant diverges at $\mathbf{s}$ for ideal zero-thickness filaments, we follow the well-established approach~\cite{Adachi-2010-PRB} and compute the induced velocity as the sum of a local contribution and a non-local contribution:
\begin{equation}
\mathbf{v}_{in}(\mathbf{s})=\beta_l\mathbf{s}'\times\mathbf{s}''+\frac{\kappa}{4\pi}\int_{\mathbf{s_1}\neq\mathbf{s}}\frac{(\mathbf{s_1}-\mathbf{s})\times d\mathbf{s_1}}{|\mathbf{s_1}-\mathbf{s}|^3},
\label{Eq-S3}
\end{equation}
where the prime denotes the derivative with respect to the arc length of the vortex filament at $\mathbf{s}$ (i.e., $\mathbf{s}'$ is the unit tangent vector along the filament, and $\mathbf{s}''$ is the unit vector along the binormal direction divided by the local curvature radius~\cite{Schwarz-1988-PRB}). The non-local term accounts for contributions from the rest of the filament and all other vortices. The coefficient $\beta_l$ is given by~\cite{Schwarz-1988-PRB}:
\begin{equation}
\beta_l=\frac{\kappa}{4\pi}\ln\left(\frac{2(l_+l_-)^{1/2}}{e^{1/4}a_0}\right)
\end{equation}
where $l_+$ and $l_-$ are the distances from point $\mathbf{s}$ to its two nearest neighbouring points along the same filament, and the cut-off parameter $a_0\simeq1$~{\AA} denotes the vortex core radius in He II.

In our simulations, the background superfluid velocity is set to $\mathbf{v}_0=0$. The time evolution of the vortex configuration is obtained by integrating Eq.~(\ref{Eq-S1}) using the fourth-order Runge-Kutta method~\cite{Press-1992-book} with a time step of $\Delta t=10^{-4}$ s. We have verified that our spatial and temporal resolutions are sufficient to ensure convergence and independence of the simulation results. During vortex evolution, whenever two vortex filaments approach each other with a minimum separation less than $\Delta\xi_{min}$, we perform a reconnection of the two filaments at the location of the minimum separation, following procedures detailed in Ref.~\cite{Tsubota-2000-PRB,Baggaley-2012-JLTP}. Furthermore, to maintain a more uniform spatial resolution along the filaments, we adaptively remove or insert vortex-filament points at each time step: a point is removed if the separation between two adjacent points falls below $\Delta\xi_{\text{min}}$, and inserted if it exceeds $\Delta\xi_{\text{max}}$.

At zero temperature, Kelvin waves excited on the vortices can lead to local kinks, promoting reconnections and the formation of small vortex loops. Through nonlinear interactions, these waves cascade to smaller scales, generating increasingly smaller loops in a self-similar breakup process~\cite{Tsubota-2000-PRB}. When loops become sufficiently small, they can propagate ballistically and be absorbed by container walls. To account for this cascade-induced vortex loss, a common approach in vortex filament simulations is to remove loops shorter than a given threshold. Following Ref.~\cite{Tsubota-2000-PRB}, we remove all vortex loops with lengths less than $5\Delta\xi_{\text{min}}$ in our simulations.

\section{Isotropicity of vortex tangles}
\begin{figure}[t]
\includegraphics[width=1.0\linewidth]{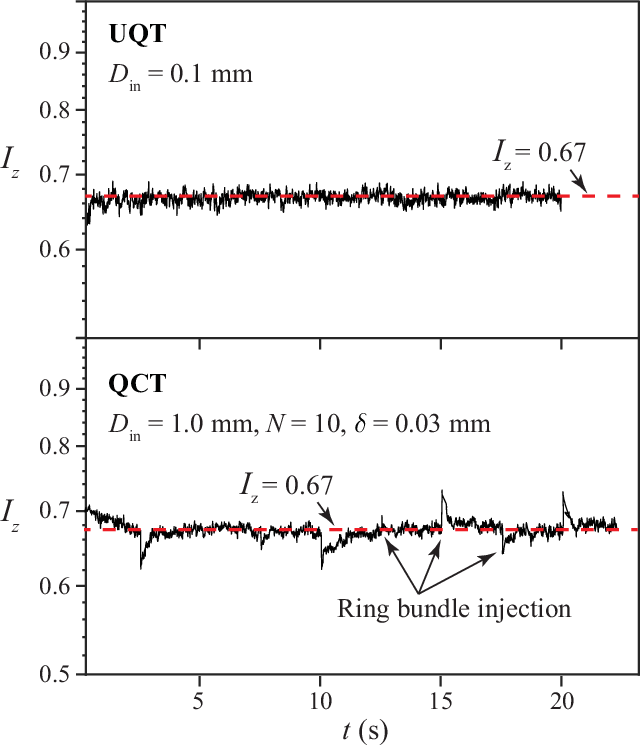}
\caption{The vortex-tangle anisotropy parameter $I_z$ as a function of time $t$ for the UQT case with $D_{\text{in}} = 0.1$ mm (top panel) and the QCT case with injection parameters $D_{\text{in}} = 1$ mm, $N = 10$, and $\delta = 0.03$ mm (bottom panel), as described in the main paper.}
\label{Suppl-Fig1}
\end{figure}
To study the dynamics of both ultra-quantum turbulence (UQT) and quasiclassical turbulence (QCT), it is essential to ensure that the generated turbulence is isotropic. This is crucial for meaningful comparisons with theoretical models developed under the assumption of isotropy. To assess whether the vortex tangles produced in our simulations meet this criterion, we evaluate the dimensionless anisotropy parameter commonly used in the literature~\cite{Adachi-2010-PRB}:
\begin{equation}
I_i=\frac{1}{\Omega \mathcal{L}}\int_{\mathcal{L}}\left[1-(\mathbf{s}'\cdot\mathbf{\hat{e}}_i)^2\right]d\xi,
\end{equation}
where $\mathcal{L}$ is the vortex-line density (i.e., the total vortex length divided by the computational volume $\Omega$), $\mathbf{s}'$ is the unit tangent vector along a vortex segment $d\xi$, and $\mathbf{\hat{e}}_i$ is the unit vector along the $i$-axis ($i = x, y, z$). The integral is performed over all vortex lines within the volume $\Omega$. For an isotropic tangle, one expects $I_x = I_y = I_z = 2/3$, whereas in the extreme case where all vortex lines are oriented perpendicular to the $i$-axis, $I_i = 1$. Fig.~\ref{Suppl-Fig1} presents the time evolution of $I_z$ for both the UQT case with $D_{\text{in}} = 0.1$ mm and the QCT case with injection parameters $D_{\text{in}} = 1$ mm, $N = 10$, and $\delta = 0.03$ mm, as described in the main paper. In both cases, $I_z$ remains close to $2/3$, indicating near-isotropic behavior. In the QCT case, small transient spikes appear in the $I_z(t)$ curve, corresponding to the injection of compact vortex ring bundles, which momentarily perturb the isotropy. The behaviors of $I_x$ and $I_y$ are found to be very similar to that of $I_z$ in both turbulence regimes, supporting the conclusion that the vortex tangles in our simulations are effectively isotropic.

\section{Energy spectrum and dissipation rate calculation}
Given the vortex configuration at time $t$, the full three-dimensional superfluid velocity field $\mathbf{v}_s$ can be computed using Eq.~(\ref{Eq-S1}). Applying a Fourier transform to this velocity field yields the turbulent kinetic energy spectrum per unit fluid mass $E(k,t)$ as a function of wavenumber $k$. The ensemble-averaged energy spectrum, $\langle E(k) \rangle$, is then obtained by averaging $E(k,t)$ over the steady-state time window shown in Fig.~1 of the main paper:
\begin{equation}
\langle E(k) \rangle = \frac{1}{t_2 - t_1} \int_{t_1}^{t_2} E(k, t) \, dt
\end{equation}
As discussed in Ref.~\cite{Yurkina-2021-LTP}, this energy spectrum can also be calculated from the vortex filament positions using the following expression:
\begin{align}
\label{Eq:E}
\langle E(k)\rangle = \frac{\kappa^2}{(2\pi)^2} \left\langle \int_0^L \int_0^L
\mathbf{s}'_j(\xi_i) \cdot \mathbf{s}'_j(\xi_j) \, \right. \notag \\
\left. \times \frac{\sin\left(k |\mathbf{s}(\xi_i) - \mathbf{s}(\xi_j)|\right)}%
{k |\mathbf{s}(\xi_i) - \mathbf{s}(\xi_j)|} \,
d\xi_i d\xi_j \right\rangle
\end{align}
where both integrals are performed over all vortex filaments. The resulting energy spectra for both UQT and QCT under various vortex-ring injection conditions are presented in Fig.~1 of the main paper.

\begin{figure}[t]
\includegraphics[width=1.0\linewidth]{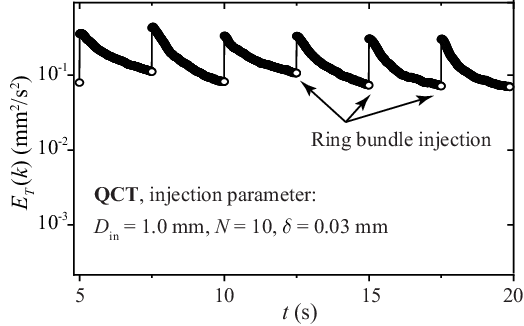}
\caption{Total turbulent kinetic energy per unit mass $E_T(t)$ as a function of time $t$ for the QCT case with $D_{\text{in}}=1$ mm, $N=10$, and $\delta=0.03$ mm.}
\label{Suppl-Fig2}
\end{figure}

In steady state, the energy dissipation rate $\epsilon$ equals the energy injection rate. To evaluate $\epsilon$, we calculate the total kinetic energy per unit mass in the computational box as $E_T(t)=\int d^3\mathbf{r} |\mathbf{u}_s(\mathbf{r})|^2=\int dk E(k, t)$. The result for the QCT case with $D_{\text{in}}=1$ mm, $N=10$, and $\delta=0.03$ mm is shown in Fig.~\ref{Suppl-Fig2}. Following each injection of a vortex ring bundle, the total energy $E_T(t)$ increases by an amount $\Delta E_T$. The energy injection rate and hence the dissipation rate is then given by $\epsilon=\Delta E_T/t_{\text{in}}$, where $t_{\text{in}}$ is the injection time interval. Averaging $\epsilon$ over all bundle injections during the steady-state time window yields $\epsilon=0.106\pm0.019$ mm$^2$/s$^3$. With $\epsilon$ determined, we can fit the energy spectrum curve $\langle E(k) \rangle$ in the wavenumber range shown in Fig.~1 of the main paper, where a clear $k^{-5/3}$ scaling is observed, using $\langle E(k)\rangle=\alpha \epsilon^{2/3}k^{-5/3}$. The fitted Kolmogorov constant is found to be $\alpha=1.43\pm0.17$.

\section{Vortex diffusion in LIA framework}
To examine whether the short-time superdiffusion of quantized vortices in both UQT and QCT is primarily governed by the local induction term—i.e., the first term on the right-hand side of Eq.~(\ref{Eq-S3})—we performed numerical simulations using the local induction approximation (LIA), in which vortex evolution is driven solely by the local term, as originally adopted by Schwarz\cite{Schwarz-1988-PRB}. It is worth noting that this analysis serves as a test case, since LIA is known to break down in capturing long-time vortex dynamics. For instance, in thermal counterflow simulations, LIA leads to the formation of artificial layered vortex structures, requiring a nonphysical mixing procedure to maintain a uniform tangle~\cite{Adachi-2010-PRB}. Moreover, LIA cannot be used to generate true QCT, as the vortex motion is governed only by local curvature effects and does not respond to large-scale flows induced by other vortices. As a result, coherent structures such as locally organized vortex bundles cannot form and maintain under LIA dynamics.

\begin{figure}[t]
\includegraphics[width=1.0\linewidth]{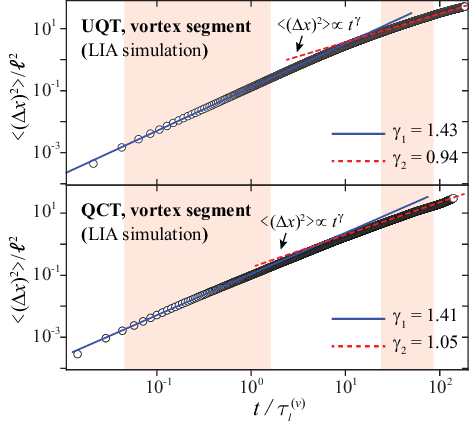}
\caption{Mean square displacement (MSD) of vortex segments, $\langle\Delta x^2(t)\rangle$, along the $x$-direction for UQT (top) and QCT (bottom), computed using the local induction approximation (LIA). Solid and dashed lines indicate power-law fits to the data within the shaded regions.}
\label{Suppl-Fig3}
\end{figure}

Nonetheless, to probe the apparent diffusion behavior of vortex segments in the absence of nonlocal interactions, we carry out LIA simulations for the UQT case with $D_{\text{in}} = 0.1$ mm, and for the QCT case with injection parameters $D_{\text{in}} = 1$ mm, $N = 1$, and $\delta = 0.03$ mm. The results are presented in Fig.~\ref{Suppl-Fig3}. The mean square displacement (MSD) of vortex segments, $\langle\Delta x^2(t)\rangle$, along the $x$-direction continues to exhibit superdiffusion scaling behavior at short times in both UQT and QCT, with the diffusion power-law exponent $\gamma_1$ slightly reduced compared to simulations using the full Biot–Savart law. This small difference reflects the contribution of the nonlocal term in the full Biot–Savart formulation. At later times, a transition to normal diffusion with a power-law exponent $\gamma_2 \simeq 1$ is observed.

\section{Vortex pair dispersion in ultra-quantum and quasiclassical turbulence}

\begin{figure*}[t]
\includegraphics[width=1.0\linewidth]{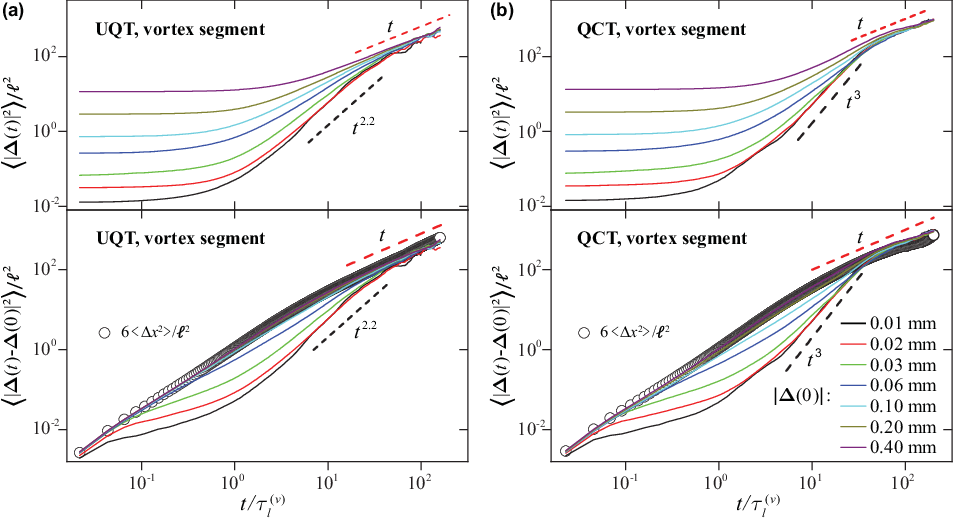}
\caption{Simulated vortex segment pair dispersion in (a) UQT and (b) QCT. The solid curves for different colors correspond to different initial pair separation $|\mathbf{\Delta}(0)|$ ranging from 0.01 mm to 0.4 mm. Open circles represent $6\langle \Delta x^2(t)\rangle/\ell^2$, obtained from diffusion data of individual vortex segments.}
\label{Suppl-Fig4}
\end{figure*}

We analyzed the two-body dispersion behavior of quantized vortices in both UQT and QCT by tracking randomly selected pairs of vortex filament points. The mean square separation, $\langle|\mathbf{\Delta}(t)|^2\rangle$, ensemble-averaged over many pairs within the steady-state time window, was evaluated for the UQT case with $D_{\text{in}} = 0.1$ mm and the QCT case with injection parameters $D_{\text{in}} = 1$ mm, $N = 1$, and $\delta = 0.03$ mm. The upper panels of Fig.~\ref{Suppl-Fig4} present the results for ensembles of vortex pairs with various initial separations $|\mathbf{\Delta}(0)|$ in both turbulence regimes.

As with the dispersion of superfluid parcels discussed in the main paper, the vortex-pair separation initially enters a “frozen” regime, where $\langle|\mathbf{\Delta}(t)|^2\rangle$ remains nearly constant due to minimal relative motion. As turbulent fluctuations accumulate, the pair separation begins to grow. In UQT, as $|\mathbf{\Delta}(0)|$ decreases, the dispersion curves collapse onto a common power-law behavior, $\langle|\mathbf{\Delta}(t)|^2\rangle \propto t^{2.2}$, similar to that observed for superfluid parcels. This scaling persists until a transition to normal diffusion occurs at later times. In QCT, for sufficiently small initial separations $|\mathbf{\Delta}(0)|$, we observe the classical Richardson scaling, $\langle|\mathbf{\Delta}(t)|^2\rangle \propto t^3$. It should be noted that we are examining the dispersion of vortex segment pairs, not passive tracer particles in a hydrodynamic flow. Therefore, quantities such as the inertial range and the turbulent kinetic energy dissipation rate $\epsilon$, which characterize the superfluid velocity field, are not applicable in this context. As a result, the classical Richardson expression, $\langle|\mathbf{\Delta}(t)|^2\rangle = g\epsilon t^3$, cannot be directly used to fit the vortex-pair dispersion curves to evaluate the Richardson constant $g$.

To exclude the influence of the initial ``frozen'' regime, we also analyzed $\langle|\mathbf{\Delta}(t) - \mathbf{\Delta}(0)|^2\rangle$ for vortex segment pairs. The results for both UQT and QCT are shown in the lower panels of Fig.~\ref{Suppl-Fig4}. For pairs with large initial separations, the normalized dispersion $\langle|\mathbf{\Delta}(t) - \mathbf{\Delta}(0)|^2\rangle / \ell^2$ closely matches the normalized diffusion data $6\langle \Delta x^2(t)\rangle / \ell^2$ for individual vortex segments. This is expected when the motions of the two segments are effectively uncorrelated. As the initial separation decreases, the curves converge and recover the dispersion scaling discussed earlier.

\bibliography{Suppl-reference}